\documentclass[11pt,twoside]{article}
\usepackage{./asp2014}

\usepackage{hyperref}
\usepackage{subfigure}
\usepackage[section]{placeins}

\aspSuppressVolSlug
\resetcounters

\begin{document}

\title{The viscous disk properties of 80 Be stars}
\author{R.~G.~Vieira$^1$ and A.~C.~Carciofi$^1$
\affil{$^1$Instituto de Astronomia, Geof\'isica e Ci\^encias Atmosf\'ericas, Universidade de S\~ao Paulo, Rua do Mat\~ao, 1226, Cidade Universit\'aria, 05508-090, S\~ao Paulo,
Brazil; \email{rodrigo.vieira@iag.usp.br}}}

\paperauthor{R.~G.~Vieira}{rodrigo.vieira@iag.usp.br}{}{Universidade de S\~ao Paulo}{Instituto de Astronomia, Geof\'isica e Ci\^encias Atmosf\'ericas}{S\~ao Paulo}{SP}{05508-090}{Brazil}
\paperauthor{A.~C.~Carciofi}{}{}{Universidade de S\~ao Paulo}{Instituto de Astronomia, Geof\'isica e Ci\^encias Atmosf\'ericas}{S\~ao Paulo}{SP}{05508-090}{Brazil}

\begin{abstract}
We investigated the properties of $80$ Be star disks, using the time dependent viscous decretion disk (VDD) model to interpret their mid-infrared continuum fluxes. Besides estimating the typical parameter ranges for such disks, this approach allowed us to relate the disk structure to its evolutionary status. In this contribution, we present a brief summary of the recently published results, and discuss some properties and time-scales of the VDD model solution.
\end{abstract}

\section{Introduction}\label{sect:intro}

Classical Be stars are characterized by Balmer lines in emission, which arise from the hot gas around these objects. The mechanism responsible for ejecting this circumstellar material acts on top of a rotation velocity close to its critical value, and remains yet unknown \citep{porter2003,rivinius2013}.

It has been almost three decades since \citet{waters1987b} determined the disk structure of $54$ classical Be stars, based on the interpretation of their \textit{IRAS} fluxes. They assumed an outflowing disk with a power-law radial density profile, $\rho(r)\propto r^{-n}$, and an opening angle of $15^{\circ}$. Among their results, they found that the exponent $n$ typically lies between $2$ and $3.5$. By adopting an outflowing radial velocity of $5\,\mathrm{km\,s^{-1}}$ at the disk base, they estimated mass loss rates ranging from $10^{-9}$ to $10^{-7}\,\mathrm{M_{\odot}\,yr^{-1}}$. These results still remain as an important benchmark in the field.

Since then, evidences favoured a rotationally supported structure rather than an outflowing disk model \citep[e.g.,][]{hanuschik1995}. In this context, the viscous decretion disk (VDD) model \citep{lee1991} emerged as the currently accepted view of Be star disks \citep{rivinius2013}. Furthermore, a much larger infrared (IR) photometric database is available now, thanks to contributions of survey missions such as \textit{ISO}, \textit{AKARI}, \textit{WISE} and \textit{Spitzer}, to mention a few.

In the light of the recent developments, we decided to revisit  \citeauthor{waters1987b}'s seminal work, and investigated the statistical properties of the disk structure of $80$ Be stars. Additionally, we took a further step and interpreted our results in terms of the time-dependent solution of the VDD model.

In this contribution, we briefly summarize the main results found in this study, and discuss some of the disk properties during build-up and dissipation. The interested reader may find the complete description of these results in \citet{vieira2016}.

\section{Summary of previous work}

To interpret the IR continuum fluxes of Be star disks, we used the pseudo-photosphere model \citep{vieira2015}. As an approximation of the VDD model, we adopt a power-law density,
\begin{equation}
\rho(r, z) = \rho_0\left(\frac{r}{R_{\mathrm{eq}}}\right)^{-n}\,\exp\left(-\frac{z^2}{2 H^2}\right),
\end{equation}
where $\rho_0$ is the base density, $R_{\mathrm{eq}}$ is the equatorial stellar radius, and $H\propto r^{1.5}$ is the disk scale height. The model assumes an isothermal and geometrically thin disk, and includes both free-free and bound-free hydrogen opacities in the LTE\footnote{Local thermodynamic equilibrium} regime \citep{brussaard1962}. The model also takes into account the stellar rotation effects, such as stellar flattening and gravity darkening. The stellar geometry is described by a Roche model \citep[e.g.,][]{cranmer1996}, and the gravity darkening exponent \citep{vonzeipel1924} was computed according to the prescription given by \citet{espinosa2011}.

From the list of B stars with stellar fundamental parameters determined by \citet{fremat2005}, we selected the classical Be stars with IR fluxes available  in at least one the missions \textit{IRAS}, \textit{AKARI/IRC} or \textit{WISE}. Cases with no IR excess (which indicates the absence of a disk) and objects with shell line profiles (for which the pseudo-photosphere model does not work; see \citealt{vieira2015} for details) were excluded. Since the missions occurred in different epochs, the SED of each mission was separately fitted. The final sample includes $80$ Be stars and $169$ SED fittings.

\begin{figure}[!t]
\begin{center}
\includegraphics[width=1. \columnwidth,angle=0]{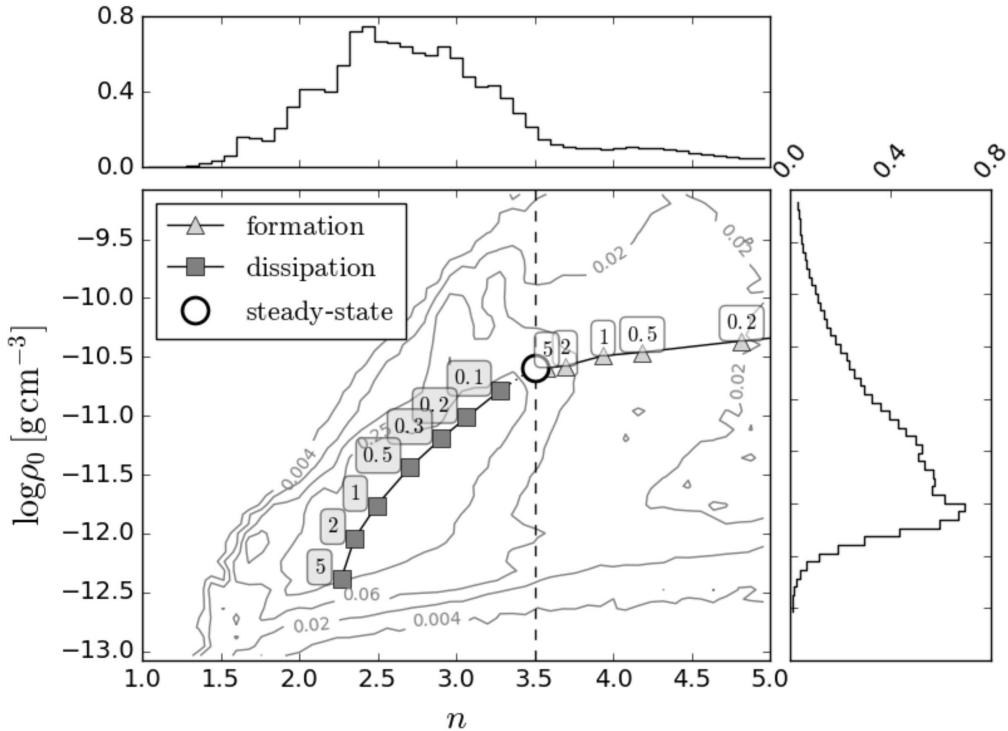}
\caption{Distribution of the $n$ and $\rho_0$ parameters, sampled with the MCMC code for the $169$ SED observations. The main panel shows the two-dimensional distribution of these parameters. The computed evolutionary track is overplotted, corresponding to a model of a B2 star, with a constant mass injection rate $\dot{M}_{\mathrm{inj}}=5\times10^{-8}\,\mathrm{M_{\odot}\,yr^{-1}}$ during disk formation, viscosity parameter $\alpha=1$ \citep{shakura1973}, and face-on orientation. The values indicated in the track correspond either to the time the disk started being fed during formation (departing from disk-less state), or the time since mass injected was turned-off, after the onset of disk dissipation. The upper and right panels correspond, respectively, to the $n$ and $\rho_0$ distributions. In all panels, the integrated distributions were normalized to unity.}
\label{fig:track}
\end{center}
\end{figure}

The disk parameters, $\rho_0$ and $n$, were constrained with a Markov chain Monte Carlo (MCMC) implementation \citep[\texttt{emcee} code\footnote{\url{http://dan.iel.fm/emcee/current/}};][]{foreman2013}. The resulting $169$ posterior probabilities sampled with the MCMC code were combined and are presented in Fig.~\ref{fig:track}. Most of the solutions lie over a dominant ridge in this distribution, and the upper left corner of the diagram is practically empty. For most of the cases, $1.5\lesssim n\lesssim 3.5$ and $-12.5\lesssim \log\rho_0\lesssim -10$.

The distribution of the disk parameters over the $n-\log\rho_0$ diagram was interpreted with a time-dependent implementation of the VDD model \citep[\texttt{SINGLEBE;}][]{okazaki2002} that  solves the one-dimensional viscous diffusion of the disk material, assuming an azimuthally symmetric mass injection at the stellar equator. To simulate an observation, we used the resulting density profiles as an input to compute the corresponding IR SED. Finally, we applied the MCMC code to fit a simple power-law model to the synthetic spectra, and determined the $n$ and $\rho_0$ values at different epochs of the disk evolution. One example of a resulting evolutionary track is shown in Fig.~\ref{fig:track}, main panel. This approach allowed us to relate specific regions of the $n-\log\rho_0$ diagram to the evolutionary status of the disk \citep{vieira2016}.

\section{Properties of the time-dependent VDD model}

\citet{vieira2016} focused on the interpretation of the IR SED during disk formation and dissipation. The flux measured at each bandpass is an integrated quantity, and takes into account the contribution of the entire disk. In this section, we restrict the discussion to the local properties of the inner disk region.

\begin{figure}[!t]
\begin{center}
\includegraphics[width=1 \columnwidth,angle=0]{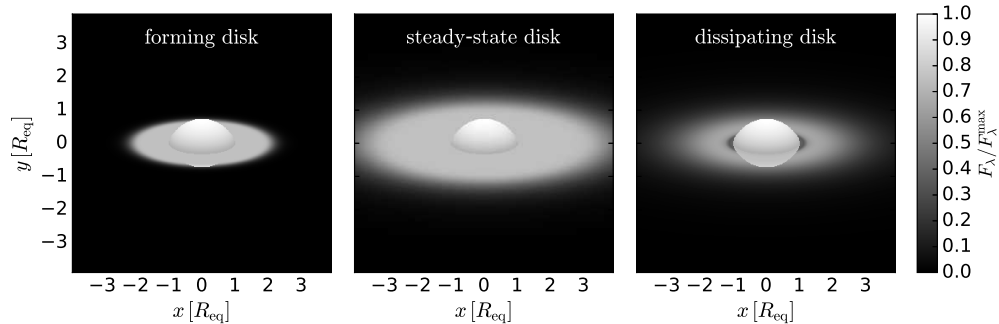}
\caption{Synthetic images of different disk dynamical states, computed at $10\,\mathrm{\mu m}$. The model parameters are the same as those adopted to compute the evolutionary track in Fig.~\ref{fig:track}, except for the disk inclination of $70^{\circ}$.}
\label{fig:synth_img}
\end{center}
\end{figure}

Figure~\ref{fig:synth_img} shows the synthetic images of a Be star at different evolutionary stages. The first panel shows a $10$~days-old disk, and exemplifies how fast the density at the inner disk grows. After $\sim$$5$~years of constant mass injection rate, the disk stops growing at $10\,\mathrm{\mu m}$, and reaches the steady-state. Strictly, the disk must be fed during an infinite time to reach the steady-state. However, mid-IR observations probe only the few stellar radii inner region of a Be disk, and is thus insensitive to further growth. Finally, when mass injection rate ceases, the disk starts to fade. The last panel in Fig.~\ref{fig:synth_img} shows a disk which dissipated for $1$~year, departing from steady-state. Notice that a gap in the intensity map appears between the star and the disk. This happens because the accretion velocities become quite high close to the star and, as a consequence, the densities become low \citep{haubois2012}.

\begin{figure}[!p]
\begin{center}
        \subfigure[Disk density at midplane.]{%
            \label{fig:sigma}
            \includegraphics[width=1.\linewidth]{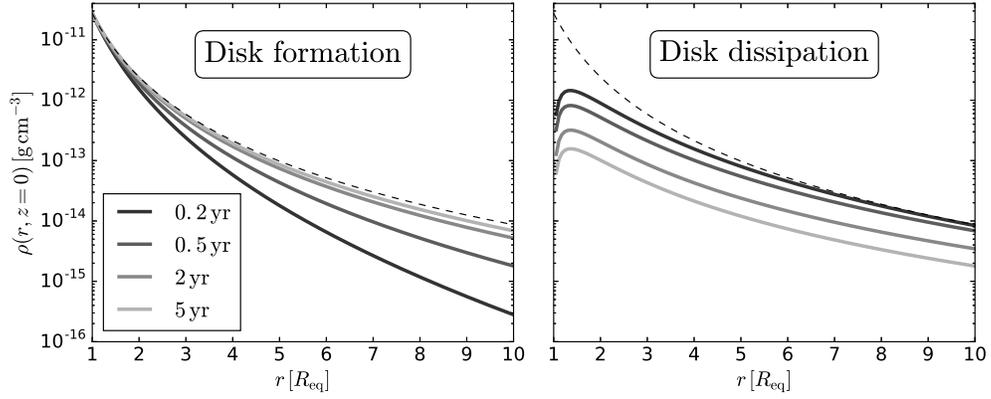}
        }\\
        \subfigure[Disk brightness profile at $10\,\mathrm{\mu m}$.]{%
            \label{fig:intens}
            \includegraphics[width=1.\linewidth]{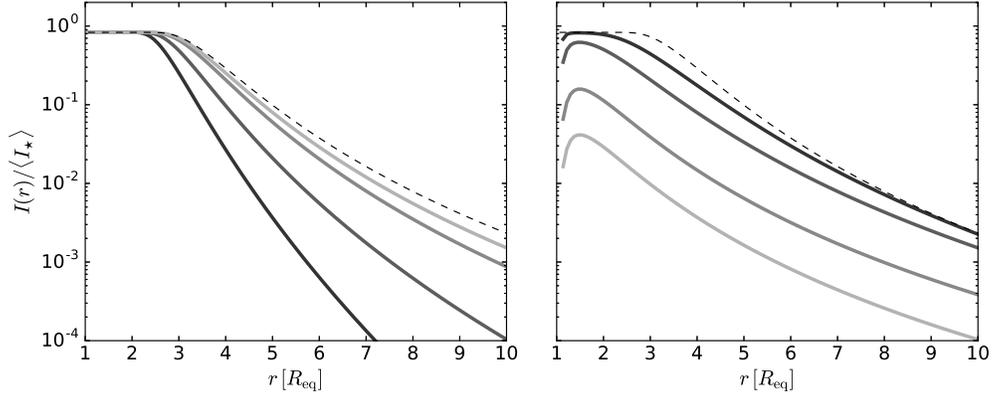}
        }\\
        \subfigure[Squared visibilities at $10\,\mathrm{\mu m}$.]{%
            \label{fig:vis}
            \includegraphics[width=1.\linewidth]{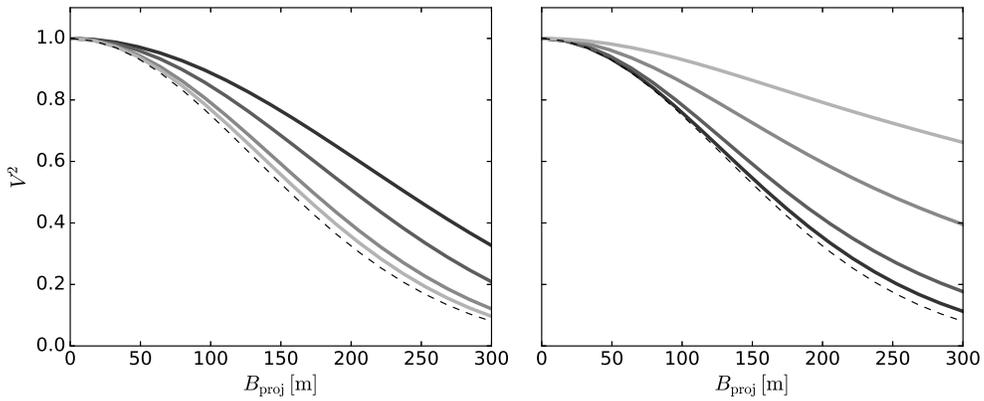}
        }\\
\end{center}
\caption{Disk evolution, as seeing from several disk properties. The left panels correspond to disk formation, while the right panels to disk dissipation. Dashed lines correspond to the steady-state solution. The adopted model parameters are the same as those used to compute the evolutionary track in Fig.~\ref{fig:track}, and the curve colors correspond either to the formation time (from disk-less state) or to the dissipation time (from steady-state).}
\label{fig:evolution}
\end{figure}

Figure~\ref{fig:evolution} shows some disk properties during formation and dissipation. In the upper panels, the disk density evolution computed with \texttt{SINGLEBE} is shown. The region inside a few stellar radii approaches the steady-state profile in less than a year. Conversely, the disk inner region re-accretes relatively quickly during dissipation, originating a low density ring at the disk base.

Using the \texttt{SINGLEBE} density profiles as an input, and adopting \citet{brussaard1962} LTE opacities, we computed the disk brightness profile for the selected dynamical stages (middle panels in Fig.~\ref{fig:evolution}).
During disk formation, we clearly see the growth of a pseudo-photosphere
As the disk grows, its ``effective'' $n$ value (i.e., average value probed by mid-IR observations) decreases, and consequently the slope of the outer part of the brightness profile becomes shallower. This is in agreement with the results found by \citet{vieira2015}, who predicted the power-law behavior $I(r)\propto r^{-2n+3/2}$ for the disk outer part.

Finally, the bottom panels in Fig.~\ref{fig:evolution} show the predictions for the squared visibilities as seeing by an interferometer at $10\,\mathrm{\mu m}$. Again, in less than one year the forming disk is very close to the steady-state. Besides, a shallower visibility profile is indicative of a dissipative disk stage. These predictions may be applied to the observations of the instrument MIDI/VLTI\footnote{Very Large Telescope Interferometer, from the European Southern Observatory (ESO)}, and to the upcoming results from the new interferometer MATISSE/VLTI under construction.

Notice the presented time-scales correspond to our specific choice for the viscosity parameter, $\alpha=1$. This parameter determines the disk viscous time scale, which goes as $1/\alpha$.

\section{Final remarks}

We presented the properties of a time-evolving viscous disk, with the corresponding typical time-scales for this evolution. When the disk structure is probed by mid-IR observations (few stellar radii), the disk reaches the steady-state appearance in only a few months, while it may take years to vanish (i.e., to become undetectable in the mid-IR). For the detailed description of the study of the $80$ Be star disks, we refer the interested reader to the recently published paper by \citet{vieira2016}.

\acknowledgements  This work made use of the computing facilities of the Laboratory of Astroinformatics (IAG/USP, NAT/Unicsul), whose purchase was made possible by the Brazilian agency FAPESP (grant No 2009/54006-4) and the INCT-A.  R.~G.~V. acknowledges the support from FAPESP (grant No 2012/20364-4). A.~C.~C acknowledges support from CNPq (grant No 307076/2012-1) and FAPESP (grant No 2015/17967-7).

\FloatBarrier
\bibliography{bib}  

\end{document}